\documentclass[twocolumn,twocolappendix]{aastex701}
\usepackage{bm,xcolor,CJK,amsmath}

\begin{document}
\begin{CJK*}{UTF8}{ipxg}
\title{Magnetic Field Amplification and Particle Acceleration \\in Weakly Magnetized Trans-relativistic Electron-ion Shocks}

\author[orcid=0000-0002-1876-5779]{Taiki Jikei (寺境太樹)}
\affiliation{Department of Astronomy and Columbia Astrophysics Laboratory, Columbia University, 538 West 120th Street, New York, NY 10027, USA}
\affiliation{Department of Earth and Planetary Science, The University of Tokyo, 7-3-1 Hongo, Tokyo 113-0033, Japan}
\email[show]{t.jikei@columbia.edu} 

\author[orcid=0000-0002-5408-3046]{Daniel {Gro{\v{s}}elj}} 
\affiliation{Centre for mathematical Plasma Astrophysics, Department of Mathematics, KU Leuven, Celestijnenlaan 200B, B-3001 Leuven, Belgium}
\email{daniel.groselj@kuleuven.be}

\author[orcid=0000-0002-1227-2754]{Lorenzo Sironi}
\affiliation{Department of Astronomy and Columbia Astrophysics Laboratory, Columbia University, 538 West 120th Street, New York, NY 10027, USA}
\affiliation{Center for Computational Astrophysics, Flatiron Institute, 162 5th Avenue, New York, NY 10010, USA}
\email{lsironi@astro.columbia.edu}

\begin{abstract}
We investigate the physics of quasi-parallel trans-relativistic shocks propagating in weakly magnetized plasmas by means of long-duration two-dimensional particle-in-cell simulations.
The structure of the shock precursor is shaped by a competition between the Bell instability and the Weibel (filamentation) instability. 
The Bell instability is dominant at relatively high magnetizations $(\sigma\gtrsim10^{-3})$, whereas the Weibel instability prevails at lower magnetizations $(\sigma\lesssim10^{-4})$.
Shocks with precursors shaped by Bell modes efficiently accelerate ions, converting a fraction $\varepsilon_{\mathrm{i}}\sim0.2$ of the upstream flow energy into downstream nonthermal ion energy.
The maximum energy of nonthermal ions exhibits a Bohm scaling in time, as $E_{\max}\propto t$.
A much smaller fraction $\varepsilon_{\mathrm{e}}\ll0.1$ of the upstream flow energy goes into downstream nonthermal electrons in the Bell regime.
On the other hand, when the precursor is dominated by Weibel modes, the shock efficiently generates both nonthermal ions and electrons with $\varepsilon_{\mathrm{i}}\sim\varepsilon_{\mathrm{e}}\sim0.1$, albeit with a slower scaling for the maximum energy, $E_{\mathrm{max}}\propto t^{1/2}$.
Our results are applicable to a wide range of trans-relativistic shocks, including the termination shocks of extragalactic jets, the late stages of gamma-ray burst afterglows, and shocks in fast blue optical transients.
\end{abstract}

\section{Introduction} \label{sec:intro}
Collisionless shocks are among the most efficient particle accelerators in the universe \citep{Drury1983, Blandford1987}.
They can be sources of cosmic rays and intense nonthermal emission.
Ultra-relativistic shocks, with shock Lorentz factors $\Gamma_{\mathrm{sh}}\gg10$, have been studied extensively in the context of gamma-ray burst (GRB) afterglow emission and ultra-high-energy cosmic ray (UHECR) production.
Theory and particle-in-cell (PIC) simulations have revealed that the Weibel filamentation instability \citep{Weibel1959,Fried1959,Medvedev1999,Silva2003} can generate intense magnetic fields in these shocks.
\footnote{
Strictly speaking, the Weibel instability (driven by temperature anisotropy) and the filamentation instability (driven by counterstreaming beams) should be treated separately \citep[e.g.,][]{Bret2009}.
Hereafter, however, we refer to the latter as the Weibel instability, following the prevailing convention in the astrophysical community.
}
PIC simulations also show efficient particle acceleration by the Fermi process \citep{Spitkovsky2008b,Sironi2013,Groselj2024}.

Late-time observations of the binary neutron star merger event GW170817 \citep{Margutti2018,Hajela2019,Hajela2022} and the recent discovery of fast blue optical transients (FBOTs), such as the AT2018cow \citep{Ho2019,Margutti2019}, direct our attention to trans-relativistic shocks, with Lorentz factors of a few.
The most important difference between trans-relativistic shocks and ultra-relativistic shocks is that trans-relativistic shocks can generally be subluminal, i.e., high-energy particles can outrun the shock and escape far upstream along magnetic field lines \citep{Sironi2009}.
This allows a potential amplification of the upstream magnetic field via the Bell streaming instability, similar to non-relativistic shocks \citep{Bell2004,Caprioli2014b,Park2015}.
At the same time, the mean energy per particle of trans-relativistic shocks is much larger than their non-relativistic counterparts, which makes them compelling candidates for UHECR production and bright synchrotron emission.

\citet{Crumley2019} performed PIC simulations of trans-relativistic shocks with $\sigma\gtrsim10^{-3}$, where $\sigma$ is the ratio of the ambient magnetic field energy to the plasma rest-mass energy (see Section \ref{sec:method}).
They show that Bell instability can indeed be triggered in the upstream of trans-relativistic shocks, and those shocks are efficient hadronic accelerators.
Their results are applicable to trans-relativistic shocks with moderate upstream magnetization, e.g., the termination shocks of active galactic nuclei (AGN) jets and microquasars.
However, late-time GRB afterglows and FBOTs propagate into weakly magnetized media, e.g., a typical interstellar-like medium has a magnetization of $\sigma\sim10^{-9}$.
The shock micro-physics in this parameter range could be qualitatively different from that at higher magnetizations.
Most notably, the fraction of energy transferred to nonthermal electrons, in \citet{Crumley2019}, seems too small to explain the bright X-ray emission from the trans-relativistic stage of GW170817 \citep{Hajela2019}.

In this paper, we investigate the physics of weakly magnetized trans-relativistic shocks with long-duration PIC simulations.
We show that there is a transition from a Bell-dominated regime at high magnetizations $(\sigma\gtrsim10^{-3})$, to a Weibel-dominated regime at low magnetizations $(\sigma\lesssim10^{-4})$.
These plasma instabilities could also play a role at the shock front and in the downstream region, but we focus on those excited in the upstream region (see Section \ref{sec:results} for details).
Ions are accelerated more efficiently, as compared to electrons, in Bell-dominated cases.
On the other hand, Weibel-dominated cases exhibit significantly larger nonthermal electron energy fractions.
Our results are primarily applicable to the late stages of GRB afterglows and FBOTs, but we also comment on the implications of our work for the termination shocks of relativistic jets.

\section{Method} \label{sec:method}
\begin{figure}[ht!]
\includegraphics[width=\columnwidth]{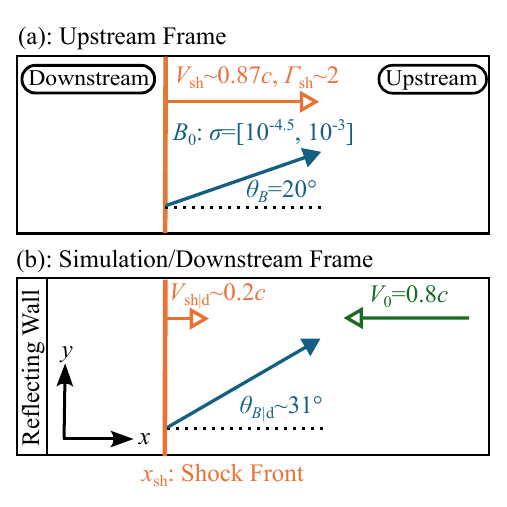}
\caption{Schematics of the setup. Panel (a) corresponds to the upstream frame, in which $\sigma$ and $\theta_B$ are defined. Panel (b) is the downstream frame, in which the PIC simulations are performed. We work in the $x-y$ plane with an in-plane background magnetic field.}
\label{fig:setup}
\end{figure}

We perform 2D PIC simulations with the OSIRIS code \citep{Fonseca2002,Fonseca2013}.
Figure \ref{fig:setup} shows the schematics of the configuration.
Panel (a) corresponds to the upstream frame, in which the following physical quantities that characterize a collisionless shock are defined.
The shock velocity $V_{\mathrm{sh}}$ is the speed at which the shock front moves in the upstream.
We set up the simulations so that the expected shock Lorentz factor in the upstream frame is $\Gamma_{\mathrm{sh}}=1/\sqrt{1-(V_{\mathrm{sh}}/c)^2}\sim2$.
The angle $\theta_B$ is the angle between the ambient magnetic field direction and the shock normal.
We focus on quasi-parallel shocks with $\theta_B=20^{\circ}$.
Finally, the shock magnetization $\sigma$ is defined here in the upstream frame as: 
\begin{equation}
\sigma=\frac{B^2_0}{4\pi n_0(m_{\mathrm{i}}+m_{\mathrm{e}})c^2},
\end{equation}
where $B_0$ is the amplitude of the ambient magnetic field, $n_0$ is the number density of the upstream ions or electrons, and $m_s$ is the mass of particle species $s$. 
In this paper, we investigate the physics of weakly magnetized shocks with the magnetization ranging from $\sigma=10^{-4.5}$ to $\sigma=10^{-3}$.
The unmagnetized case $(\sigma=0)$ is discussed in Appendix \ref{app:unmagnetized}.

Figure \ref{fig:setup}(b) shows the simulation setup.
We work in the $x-y$ plane, in which the shock propagates in the positive $x$-direction.
The upstream plasma with velocity $V_0=-0.8c$ collides into the reflecting wall located at $x=0$.
The simulation frame approximately coincides with the downstream frame.
Strictly speaking, the downstream plasma has a finite drift velocity in the $y$-direction, but such a velocity is much smaller than the shock velocity.
The shock propagates with $V_{\mathrm{sh|d}}\sim0.2c$, resulting in the target $\Gamma_{\mathrm{sh}}\sim2$.
The inclination of the ambient magnetic field in the simulation frame is $\theta_{B\mathrm{|d}}=\mathrm{atan}(\gamma_0\mathrm{tan}\theta_B)\sim31^{\circ}$ due to the Lorentz transformation, where $\gamma_0=1/\sqrt{1-(V_0/c)^2}$.
We mark the location of the shock front as $x=x_{\mathrm{sh}}$.

Our simulation box has a $y$-size of $L_y=163.84\,d_{\mathrm{i}}$, where $d_{\mathrm{i}}=c/\omega_{\mathrm{pi}}$ is the ion skin depth, and $\omega_{\mathrm{p}s}=(4\pi n_0e^2/m_s)^{1/2}$ is the plasma frequency of species $s$.
A periodic boundary condition is used in the $y$-direction.
We use a moving injector that moves in the positive $x$-direction at the speed of light to continuously supply the upstream plasma.
We use a cell size of $\Delta x=0.4\,d_{\mathrm{e}}$, where $d_{\mathrm{e}}=c/\omega_{\mathrm{pe}}$ is the electron skin depth, and the time step is $\Delta t=0.5\Delta x/c$.
The upstream number of particles per cell is $N_{\mathrm{ppc}}=16$ per species, with cubic spline shapes.
To save computational time, we use a reduced ion-to-electron mass ratio of $m_{\mathrm{i}}/m_{\mathrm{e}}=100$.

We made the following optimizations to minimize any numerical artifacts.
An alternative stencil to the standard Yee stencil \citep{Blinne2018} is used to suppress the numerical Cherenkov instability \citep[see][]{Groselj2022}.
We apply 16 passes, in each direction, of binomial filtering to the electric current at each time step to further reduce the noise.
These strategies allowed us to evolve the system for an unprecedentedly long time of $\omega_{\mathrm{pi}}t=7000$, and up to $\omega_{\mathrm{pi}}t=12000$ in some cases.
We initialize the upstream plasma with a velocity profile that is zero at the $x=0$ boundary, and linearly transitions to the upstream value $V_0$ at $x=100\,d_{\mathrm{i}}$ \citep{Groselj2024}.
The choice of initial flow profile suppresses the unphysically strong initial reflection of particles from the left wall before the shock is formed.

\section{Results} \label{sec:results}
\subsection{Shock Structure} \label{subsec:shock_structure}
\begin{figure}[ht!]
\includegraphics[width=\columnwidth]{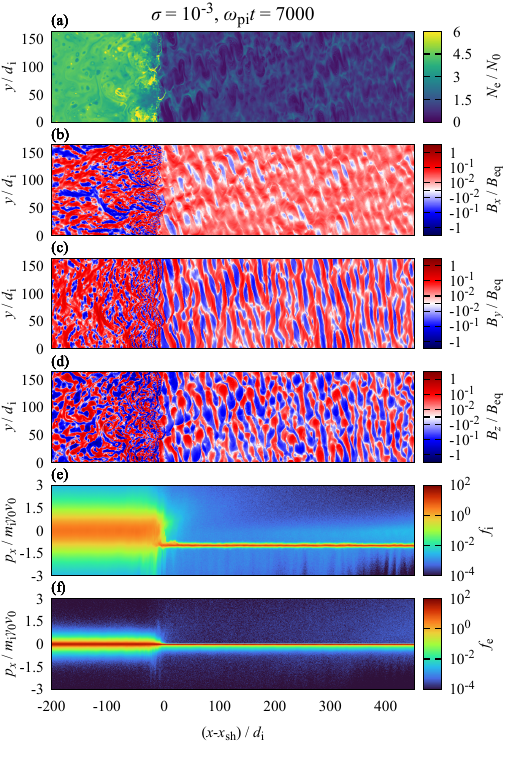}
\caption{Snapshot of the $\sigma=10^{-3}$ shock, taken at $\omega_{\mathrm{pi}}t=7000$. 
Panel (a) is the plasma density $N_{\mathrm{e}}$ normalized by the upstream value. 
Panels (b-d) are the magnetic field components in units of the equipartition magnetic field, Equation (\ref{eq:b_eq}). 
The colorbars are in a symmetric log scale, in which the range $[-0.01, 0.01]$ is in a linear scale, and values outside this range are in a log scale. 
Panels (e, f) are the phase space densities $f(x, p_x)$ of ions and electrons, respectively.
(An animation of this figure is available in the journal version.)} 
\label{fig:1e-3}
\end{figure}

We compare the shock structures for different magnetizations.
Figure \ref{fig:1e-3} shows a snapshot of the $\sigma=10^{-3}$ case, taken at $\omega_{\mathrm{pi}}t=7000$.
Panel (a) shows the simulation frame electron density $N_{\mathrm{e}}$, normalized by the far upstream value $N_0=n_0\gamma_0$.
The position of the shock front $x_{\mathrm{sh}}$ is defined as the point where the $y$-averaged $N_{\mathrm{e}}=2.5N_0$. 
Panels (b-d) show the magnetic field components, normalized by the equipartition field strength:
\begin{equation} \label{eq:b_eq}
B_{\mathrm{eq}}=[4\pi n_0\gamma_0^2(m_{\mathrm{i}}+m_{\mathrm{e}})c^2]^{1/2}.
\end{equation}

Interaction between the cosmic rays and the incoming upstream plasma results in a structure typical of the late stage of the Bell instability \citep{Caprioli2014b, Crumley2019}.
In this paper, we define cosmic rays as nonthermal particles that are reflected and energized at the shock and propagate into the upstream.
The quantitative definition and identification method of these cosmic rays in the simulations are discussed in Subsections \ref{subsec:cr_current} and \ref{subsec:acceleration}.
The fluctuations, particularly in the $B_y$ component (panel (c)), have a wavevector predominantly parallel to the ambient field, with a typical spatial scale of a few tens of ion skin depths, as expected for the Bell instability \citep{Bell2004,Amato2009}.
Locally, the fields reach near equipartition strength.
The late nonlinear stage, which we are looking at, displays a large-scale structure of rarified cavities and dense filaments
(panel (a)) and $B_z$ (panel (d)) \citep{Caprioli2014a,Caprioli2014b}.
Since the Bell modes, in their linear stage, have a circular polarization, the $B_z$ structure is similar to $B_y$, modulo this nonlinear density structure.

These results are consistent with \citet{Crumley2019}, in which the parameters in their fiducial run correspond to $\Gamma_{\mathrm{sh}}\sim1.8, \theta_B=10^{\circ}$, and $\sigma=3\times10^{-3}$, respectively, with our definitions, which is relatively similar to the run in Figure \ref{fig:1e-3}.
Their simulation ends at $\omega_{\mathrm{pi}}t\sim4000$.
Since we have run the simulation longer than they have, the nonlinear filamentary structure is more pronounced.

Panels (e,f) are the phase space densities $x-p_x$ of the two species, in which $p_x$ is normalized by the momentum of the upstream ions $m_{\mathrm{i}}\gamma_0v_0$.
We can see that there is a significant amount of high-energy ions, with $|p_x|/m_{\mathrm{i}}\gamma_0v_0\gg1$, in the downstream.
However, there is a much smaller number of electrons in the same energy range.
This implies that the particle acceleration efficiency is different between the two species.
We will discuss particle acceleration in Subsection \ref{subsec:acceleration}.

\begin{figure}[ht!]
\includegraphics[width=\columnwidth]{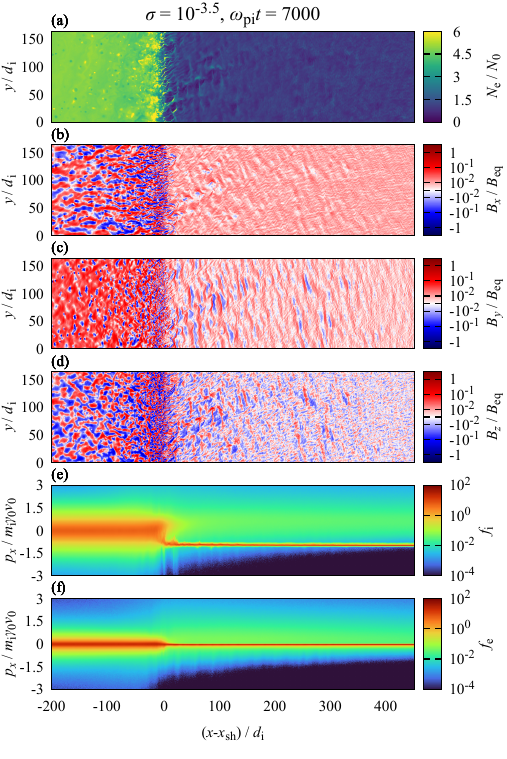}
\caption{Snapshot of the $\sigma=10^{-3.5}$ shock, taken at $\omega_{\mathrm{pi}}t=7000$. The format is the same as in Figure \ref{fig:1e-3}.
(An animation of this figure is available in the journal version.)}
\label{fig:1e-3.5}
\end{figure}

Figure \ref{fig:1e-3.5} shows the $\sigma=10^{-3.5}$ shock at the same time in the simulation $(\omega_{\mathrm{pi}}t=7000)$.
The fluctuations that dominate in the upstream $B_y$ component, panel (c), still have a wave vector parallel to the ambient field, similarly to the $\sigma=10^{-3}$ case, but their amplitude is much smaller.
This can be attributed to the cosmic ray current satisfying the \textit{high-current} condition \citep{Weidl2019}.
The high-current condition is satisfied when the cosmic ray current $J_{\mathrm{CR}}$ in the upstream becomes larger than a critical value, which depends on the upstream magnetization.
We will discuss this in Subsection \ref{subsec:cr_current}.
We can see in Figure \ref{fig:1e-3.5} (e) that there are more reflected ions in the $\sigma=10^{-3.5}$ case, compared to the $\sigma=10^{-3}$ case (Figure \ref{fig:1e-3} (e)).
The difference in the number of downstream high-energy particles between ions and electrons is less significant than in the $\sigma=10^{-3}$ case.

\begin{figure}[ht!]
\includegraphics[width=\columnwidth]{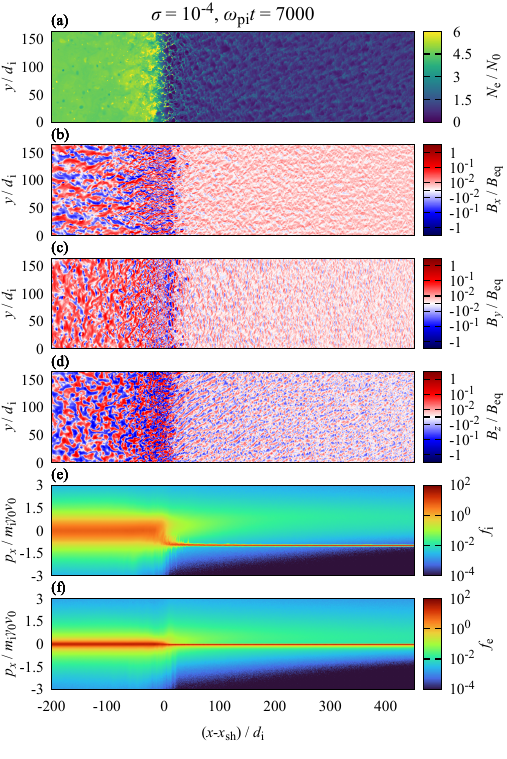}
\caption{Snapshot of the $\sigma=10^{-4}$ shock, taken at $\omega_{\mathrm{pi}}t=7000$. The format is the same as in Figure \ref{fig:1e-3}. Compared to Figure \ref{fig:1e-3}, the shock structure is notably changed, as the precursor is now shaped by the Weibel instability.
(An animation of this figure is available in the journal version.)}
\label{fig:1e-4}
\end{figure}

Finally, Figure \ref{fig:1e-4} is the $\sigma=10^{-4}$ shock at $\omega_{\mathrm{pi}}t=7000$.
The magnetic field structure in this case is distinct from the previous cases.
Most of the fluctuating magnetic field energy is in the $B_z$ component, panel (d), and little in $B_y$.
The wavenumber vector is mostly perpendicular to the ambient field, i.e., $\bm{k}\perp\hat{\bm{B}}_0$, with a characteristic spatial scale of $d_{\mathrm{i}}k\sim1$ in the downstream frame.
These are the characteristics of the Weibel instability \citep{Weibel1959, Fried1959}.
Note that cosmic rays are still streaming along the ambient field, and the magnetic field could alter the structure of the Weibel modes as compared to the case of an unmagnetized medium \citep{Lemoine2014, Grassi2017}.
We discuss the unmagnetized $\sigma=0$ case in Appendix \ref{app:unmagnetized}.

Here, we have compared the structures of shocks with fixed $\Gamma_{\mathrm{sh}}\sim2$ and $\theta_B=20^{\circ}$, varying the upstream $\sigma$.
We have shown that the dominant physics in the upstream has a transition from the Bell instability at high magnetizations $\sigma\gtrsim10^{-3}$ to the Weibel instability at lower magnetizations $\sigma\lesssim10^{-4}$.
In between, at $\sigma=10^{-3.5}$ we see a magnetic field structure, morphologically similar to Bell modes, but having a much smaller amplitude.
In Subsection \ref{subsec:cr_current}, we discuss how the magnetization and cosmic ray properties determine the dominant instability.
The $\sigma=10^{-4.5}$ case is not shown here because the overall structure is essentially the same as the $\sigma=10^{-4}$ case.
Nevertheless, it will be included in the quantitative analysis of the following Subsections.

\subsection{Cosmic Ray Driven Instabilities} \label{subsec:cr_current}
Here, we discuss the plasma streaming instabilities in the upstream, driven by the returning cosmic rays.
In Subsection \ref{subsec:shock_structure}, we discussed how the precursor is shaped by a competition between the Bell instability and the Weibel instability. 
It has recently been shown that the characteristics of Bell instability change when the current carried by the cosmic rays exceeds a certain threshold \citep{Weidl2019}.
This is called the high-current regime.
\citet{Lichko2025} suggests that the fluctuating fields $\delta B$ generated by Bell instability in the high-current regime saturate at a fraction of the ambient magnetic field energy.
In their work, the typical saturation value was $\delta B/B_0\sim5$.
We find a similar value of $\delta B/B_0\sim10$ for our parameters (see Appendix \ref{app:saturation}).
This is different from the classical low-current regime of Bell instability, in which a fixed fraction $\sim0.1$ of the cosmic ray momentum flux is converted into the amplified fields \citep{Caprioli2014b, Zacharegkas2024}.
The saturation energy density of the high-current Bell mode scales with $B_0^2$, while that of the low-current Bell mode scales with the cosmic ray momentum flux.
It is essential to distinguish these two regimes, especially because we need to understand their competition with the Weibel instability.
The high-current condition is satisfied when
\begin{equation} \label{eq:high_current}
\eta>2\sigma^{1/2}=\eta_{\mathrm{crit}},
\end{equation}
where $\eta=J_{\mathrm{CR}}/en_0c$ and $J_{\mathrm{CR}}$ is the cosmic ray current.
This is equivalent to the growth rate of the classical Bell instability becoming larger than the gyro-frequency of upstream incoming ions.

To investigate the nature of Bell mode saturation, we explicitly check Equation (\ref{eq:high_current}) in our PIC simulations.
To this end, we define cosmic rays as the ions in the upstream region that have changed the sign of their $x$-momentum, at least once, i.e., that have been reflected back by the shock \citep{Lemoine2019, Vanthieghem2022}.
We measure the cosmic ray ion four current $J^{\mu}_{\mathrm{CR}}$ at $[x-x_{\mathrm{sh}}(t)]/d_{\mathrm{i}}=[200, 300]$ in the simulation frame.
Then we compute the upstream cosmic ray current along the background magnetic field as $J_{\mathrm{CR}}=J^x_{\mathrm{CR}}|_{\mathrm{u}}/\cos{\theta_B}$, where $|_{\mathrm{u}}$ indicates quantities measured in the upstream frame, and $\theta_B$ is defined in the upstream frame.
In addition, we used the $y$ component and computed $J^y_{\mathrm{CR}}|_{\mathrm{u}}/\sin\theta_B$.
The two methods give consistent results in this region.
When the current is measured closer to the shock front, the two estimates are less consistent due to the change in the background plasma profile, e.g., resulting from the deceleration of the upstream flow \citep{Vanthieghem2022}.
We note that we have assumed that the cosmic rays are dominated by ions.
This is a common choice for Bell instability in non-relativistic shocks \citep{Bell2004,Amato2009}.
We confirm that this is also valid in trans-relativistic shocks, in Subsection \ref{subsec:acceleration}.
The role of electron cosmic rays is discussed in Appendix \ref{app:self_regulation}.

\begin{figure}[ht!]
\includegraphics[width=\columnwidth]{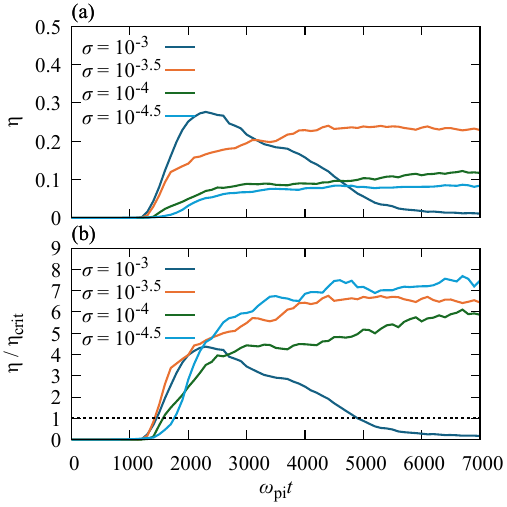}
\caption{Time evolution of the upstream cosmic ray current density for various $\sigma$. Panel (a) shows the raw $\eta=J_{\mathrm{CR}}/en_0c$, and panel (b) the normalized $\eta/\eta_{\mathrm{crit}}$. Dark teal, orange, dark green, and turquoise curves represent $\sigma=10^{-3}, 10^{-3.5}, 10^{-4}$, and $10^{-4.5}$, respectively. Cosmic ray current is measured using only ions.}
\label{fig:cr_current}
\end{figure}

Figure \ref{fig:cr_current} shows the time evolution of the cosmic ray current averaged in the spatial interval $[x-x_{\mathrm{sh}}(t)]/d_{\mathrm{i}}=[200, 300]$.
Panel (a) shows the raw $\eta$, whereas panel (b) corresponds to $\eta/\eta_{\mathrm{crit}}$.
In panel (b), we can see that the $\sigma=10^{-3}$ shock eventually self-regulates the current to a subcritical value, consistent with the strong low-current Bell magnetic fluctuations in Figure \ref{fig:1e-3}.
On the other hand, the cases with magnetization below $\sigma=10^{-3.5}$ remain supercritical, at least up to $\omega_{\mathrm{pi}}t=7000$.
Looking at $1000\lesssim\omega_{\mathrm{pi}}t\lesssim2000$ in panel (a), we notice that the two high-$\sigma$ cases are clearly distinct from the two low-$\sigma$ cases.
This is presumably because of the difference in the ion reflection process.
The classical reflection mechanism related to shock potential and mirror force \citep{Balogh2013, Burgess2015} could apply to the high-$\sigma$ cases.
On the other hand, the reflection in the low-$\sigma$ cases is governed by interaction with the small-scale Weibel fields in the vicinity of the shock front \citep{Spitkovsky2008a, Kato2008, Jikei2024b,Parsons2024}.
This results in the $\sigma=10^{-3.5}$ case, Figure \ref{fig:1e-3.5}, being more supercritical, i.e., larger $\eta/\eta_{\mathrm{crit}}$, than the $\sigma=10^{-4}$ case, despite the larger $\eta_{\mathrm{crit}}$.
Below $\sigma=10^{-4}$, the cosmic ray current is almost independent of $\sigma$; therefore, lower magnetization shocks would be more supercritical.

\begin{figure}[ht!]
\includegraphics[width=\columnwidth]{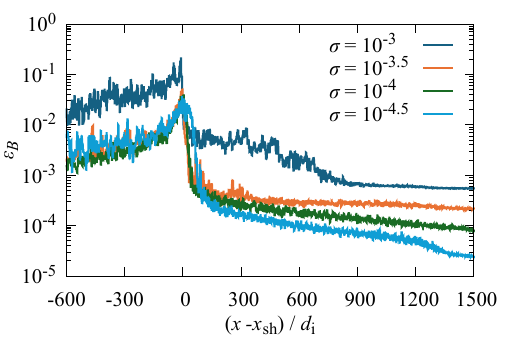}
\caption{Magnetic field energy density $\varepsilon_B$ at $\omega_{\mathrm{pi}}t=7000$ for various $\sigma$. Color scheme is the same as in Figure \ref{fig:cr_current}.}
\label{fig:eps_b}
\end{figure}

Figure \ref{fig:eps_b} shows the transversely averaged spatial profile of the magnetic energy fraction:
\begin{equation}
\varepsilon_B=\frac{B^2}{8\pi n_0\gamma_0(\gamma_0-1)(m_{\mathrm{i}}+m_{\mathrm{e}})c^2}.
\end{equation}
This corresponds to the simulation frame magnetic field energy density divided by the upstream kinetic energy density.
Note that this is a frame-dependent quantity.
The strength of the near upstream Weibel field seen at $(x-x_{\mathrm{sh}})/d_{\mathrm{i}}=[200, 500]$, in the dark green and turquoise curves is $\varepsilon_B\sim2\times10^{-4}$.
The shock transition region of the $\sigma=10^{-4.5}$ case, as seen from the $\varepsilon_B$ profile, is slightly broader, compared to the $\sigma=10^{-4}$ case.
This is due to the larger-scale structures generated during the strongly nonlinear stage of the Weibel shock evolution (see Appendix \ref{app:unmagnetized} and \citet{Groselj2024}).
For the $\sigma=10^{-3.5}$ case in the orange curve, we see some spikes peaking at $\varepsilon_B\sim10^{-3}$ at $(x-x_{\mathrm{sh}})/d_{\mathrm{i}}=[200, 300]$, which is what we observed in Figure \ref{fig:1e-3.5}(c).
The upstream magnetic field of the $\sigma=10^{-3}$ case is much stronger at $\varepsilon_B\sim10^{-2}$.
The high-current Bell modes generated in the $\sigma=10^{-3.5}$ case saturate an order of magnitude below the low-current Bell modes of the $\sigma=10^{-3}$ case, but still remain locally larger than the Weibel modes, which dominate at lower magnetizations.
In all cases, $\varepsilon_B$ approaches the far upstream value;
\begin{equation}
\varepsilon_{B,0}=\frac{\sigma}{2}\frac{\gamma_0^2\sin^2\theta_B+\cos^2\theta_B}{\gamma_0(\gamma_0-1)}\simeq0.54\,\sigma,
\end{equation}
at around $(x-x_{\mathrm{sh}})/d_{\mathrm{i}}\sim1500$.

In this Subsection, we analyzed the cosmic ray currents and the plasma instabilities driven by cosmic rays in the upstream region.
By considering the Weibel instability and both low-current and high-current Bell instability, we were able to elucidate the $\sigma$-dependence on the shock structure seen in Subsection \ref{subsec:shock_structure}.
For additional details regarding the saturation of the high-current Bell modes, see Appendix \ref{app:saturation}.

\subsection{Particle Acceleration} \label{subsec:acceleration}
\begin{figure*}[ht!]
\centering
\includegraphics[width=1.5\columnwidth]{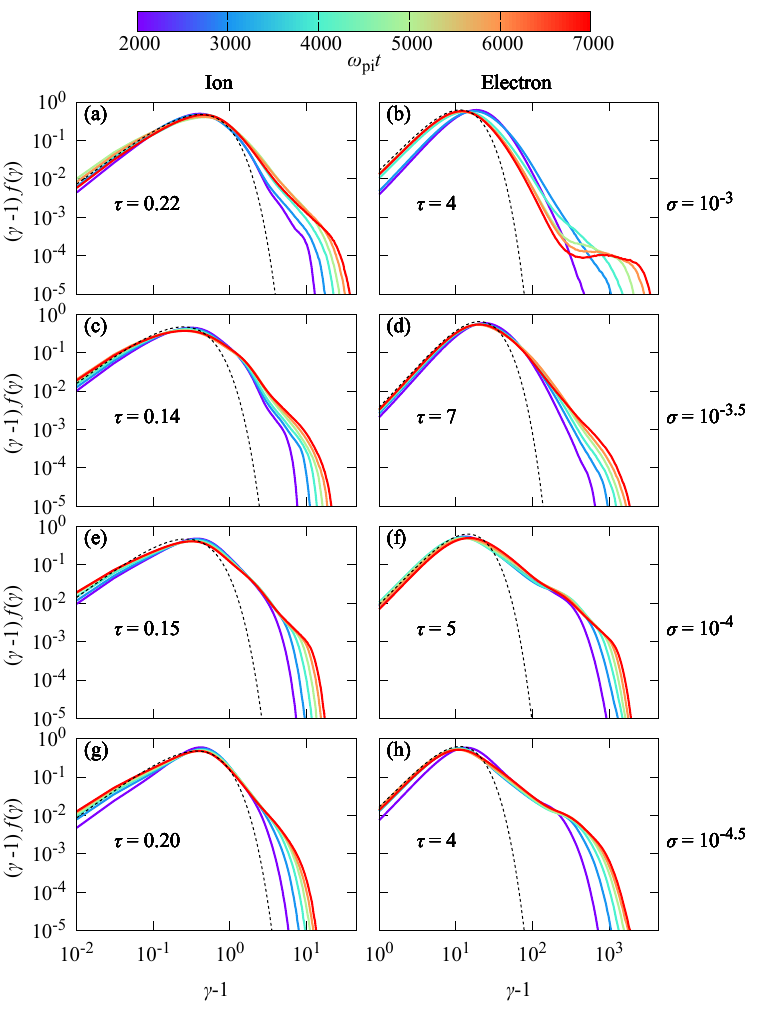}
\caption{Particle energy spectra in the downstream region. Top panels (a,b) are ions and electrons for the $\sigma=10^{-3}$ case, and bottom panels (g,h) are ions and electrons for the $\sigma=10^{-4.5}$ case, respectively. Different colors represent different times, whereas the black dotted curves are the Maxwell-J{\"u}ttner distribution with the best-fit temperature at $\omega_{\mathrm{pi}}t=7000$.}
\label{fig:acceleration}
\end{figure*}

We now discuss how particles are accelerated in these various shock regimes.
Figure \ref{fig:acceleration} shows the particle energy spectrum $(\gamma-1) f(\gamma)$ in the downstream region $(x-x_{\mathrm{sh}})/d_{\mathrm{i}}=[-200, -100]$.
The normalization is such that $\int_1^{\infty} fd\gamma=1$.
The left (a,c,e,g) and right (b,d,f,h) columns represent ions and electrons, respectively.
The top (a,b) row shows the $\sigma=10^{-3}$ case, and the magnetization decreases moving down to the bottom row with $\sigma=10^{-4.5}$.
Different colors represent simulation time, ranging from $\omega_{\mathrm{pi}}t=2000$ to $7000$ as indicated by the colorbar.
The data is overlayed with the best-fit Maxwell-J{\"u}ttner distribution for $\omega_{\mathrm{pi}}t=7000$ (black dotted curve).
Its normalized temperature is defined using the mass of each species as $\tau=k_{\mathrm{B}}T/m_sc^2$.
We find that the ion nonthermal tails keep extending in time for all magnetizations (panels (a,c,e,g)).

For electrons, the time evolution of the nonthermal spectra has a clear dependence on $\sigma$.
For the low-$\sigma$ cases (panels (f,h)), the nonthermal tail has a relatively steady normalization, and its upper cutoff keeps growing.
On the other hand, for the Bell-dominated case with $\sigma=10^{-3}$ (panel (b)),
the normalization of the tail gets suppressed after $\omega_{\mathrm{pi}}t\sim4000$.
This is because the Bell instability generates a strong magnetic field perpendicular to the shock normal, which then changes the local magnetic field angle \citep{Crumley2019}.
Although the shock started subluminal, this effect can make it superluminal for the low-energy electrons that have a Larmor radius smaller than the half-wavelength of Bell modes.
Thus, electron injection is suppressed.
The filamentary structure of nonlinear Bell modes still allows a fraction of electrons to be injected, but a significant decrease in the injection rate is inevitable, due to the superluminality constraint mentioned above.
Ions could still experience efficient injection because they have a much larger Larmor radius, even for the low-energy population at energies just above the thermal peak.
High-energy electrons, which already had a Larmor radius larger than the Bell scale, can still be accelerated.

\begin{figure}[ht!]
\includegraphics[width=\columnwidth]{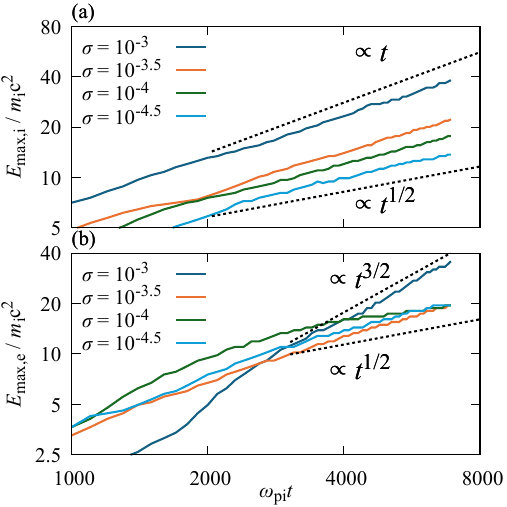}
\caption{Time evolution of the maximum energy for various magnetizations. Note that we employ a log scale on both axes. Panel (a) is for ions, (b) is for electrons. The color scheme for the magnetization is the same as in Figure \ref{fig:cr_current}. The black dotted lines represent power-law scalings.}
\label{fig:max_ene}
\end{figure}

Figure \ref{fig:max_ene} shows the time evolution of the maximum particle energy $E_{\mathrm{max},s}=m_sc^2(\gamma_{\mathrm{max}}-1)$, where $\gamma_{\mathrm{max}}$ is the Lorentz factor at which $(\gamma-1)f(\gamma)$ becomes $10^5$ times smaller than the peak.
Panel (a) shows the $E_{\mathrm{max,i}}(t)/m_{\mathrm{i}}c^2$ of ions.  
Ions are efficiently accelerated in all cases, but the scaling of the maximum Lorentz factor is different.
The nonthermal tail is growing the fastest in the highest-$\sigma$ case and the slowest in the lowest-$\sigma$ case.
In particular, we find a Bohm-like scaling $E_{\mathrm{max,i}}\propto t$ in the $\sigma=10^{-3}$ case (dark teal), whereas lower magnetization cases (dark green and turquoise) follow the $E_{\mathrm{max,i}}\propto t^{1/2}$ scaling.
The latter is consistent with ultra-relativistic shock simulations by \citet{Sironi2013}, who found that the maximum energy of particles accelerated in small-scale turbulence, such as the Weibel fields, scales as $E_{\mathrm{max}}\propto t^{1/2}$.
In contrast, the Bohm scaling $E_{\mathrm{max}}\propto t$ is expected for Bell-dominated shocks \citep{Drury1983,Gargate2012,Stockem2012}.

Figure \ref{fig:max_ene}(b) shows the maximum energy of electrons $(E_{\mathrm{max,e}}(t)/m_{\mathrm{i}}c^2)$.
As we have seen in Figure \ref{fig:acceleration}, the properties of the electron spectra and their dependence on $\sigma$ are qualitatively different from those of ions.
At the two lowest magnetizations $\sigma=10^{-4}$ and $10^{-4.5}$, we find the $E_{\mathrm{max}}\propto t^{1/2}$ scaling, which is the same as for ions in the same magnetization range.
For the $\sigma=10^{-3}$ case, we can see that the highest energy of the electron tail grows at a super-Bohm rate with $E_{\mathrm{max}}\propto t^{3/2}$.
Presumably, this is a transient feature related to the time-dependent electron injection and evolving precursor structure.
For the same reason, one should not interpret the slope of the electron nonthermal tail at late time at face value (Figure \ref{fig:acceleration}(b)), since when the injection will settle to a time-steady value, the slope of the energy spectrum would likely be steeper.

Finally, let us discuss the downstream energy partition between ions and electrons.
We define:
\begin{eqnarray} 
\mathcal{E}_s=\frac{m_s}{m_{\mathrm{i}}(\gamma_0-1)}\int_1^{\infty}(\gamma-1)f_s(\gamma)d\gamma, \label{eq:total_energy}\\
\varepsilon_s=\frac{m_s}{m_{\mathrm{i}}(\gamma_0-1)}\int_{\gamma_{\mathrm{inj}}}^{\infty}(\gamma-1)f_s(\gamma)d\gamma, \label{eq:nonthermal_energy}
\end{eqnarray}
calculated at $\omega_{\mathrm{pi}}t=7000$ in the downstream region $(x-x_{\mathrm{sh}})/d_{\mathrm{i}}=[-200, -100]$, same as in Figures \ref{fig:acceleration} and \ref{fig:max_ene}.
$\mathcal{E}_s$ and $\varepsilon_s$ are the total and the nonthermal kinetic energy density, respectively.
They are both normalized to the shock energy.
We define the injection Lorentz factor $\gamma_{\mathrm{inj}}$ as $(\gamma_{\mathrm{inj}}-1)=3(\gamma_{\mathrm{peak}}-1)$, where $\gamma_{\mathrm{peak}}$ is the Lorentz factor at which $(\gamma-1)f(\gamma)$ takes its maximum value.

\begin{figure}[ht!]
\includegraphics[width=\columnwidth]{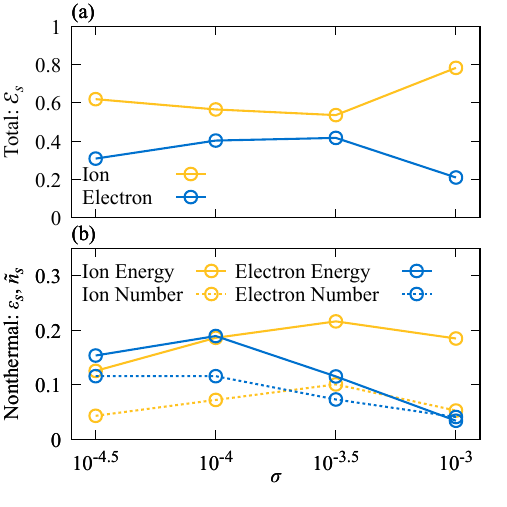}
\caption{
Magnetization dependence of the downstream kinetic energy partition and the fraction of nonthermal particles. 
Panel (a) shows the total energy density (Equation (\ref{eq:total_energy})), and panel (b) shows the nonthermal energy density (Equation (\ref{eq:nonthermal_energy}), solid lines) and nonthermal number fraction (Equation (\ref{eq:nfraction}), dotted lines), respectively, in the downstream region.
The yellow and blue lines correspond to ions and electrons, respectively.}
\label{fig:energy_partition}
\end{figure}

Figure \ref{fig:energy_partition} (a) shows the total energy density of each particle species.
The three low-magnetization cases $(\sigma\lesssim10^{-3.5})$ show $\mathcal{E}_{\mathrm{i}}\sim0.6$ and $0.3\lesssim \mathcal{E}_{\mathrm{e}}\lesssim0.4$, respectively.
Electron heating to near equipartition with ions is consistent with the result of unmagnetized shocks in both ultra-relativistic \citep{Spitkovsky2008a,Vanthieghem2022} and non-relativistic \citep{Vanthieghem2024} regimes.
The Bell-dominated $\sigma=10^{-3}$ case exhibits a noticeably different energy partition $\mathcal{E}_{\mathrm{i}}\sim0.8$ and $\mathcal{E}_{\mathrm{e}}\sim0.2$, in other words, weaker electron heating.

The energy density of the nonthermal component is shown in Figure \ref{fig:energy_partition}(b) solid lines.
For all four magnetizations, the ion nonthermal energy is $0.1\lesssim \varepsilon_{\mathrm{i}}\lesssim0.2$, consistent with previous simulations in a wide range of $\Gamma_{\mathrm{sh}}$ and $\sigma$ \citep[e.g.,][]{Sironi2013,Caprioli2014a,Crumley2019}.
Note that the precise values of $\varepsilon_{\mathrm{i}}$ and $\varepsilon_{\mathrm{e}}$ depend on our choice of $\gamma_{\mathrm{inj}}$.
Therefore, we shall mostly focus on the dependence on $\sigma$, which is rather weak for the ion species.

On the other hand, the electron nonthermal energy has a clear dependence on magnetization.
For the two Weibel-dominated cases $(\sigma=10^{-4.5}$ and $10^{-4})$,  
$\varepsilon_{\mathrm{e}}\sim0.2$, which is similar to the ion nonthermal energy.
The value of $\varepsilon_{\mathrm{e}}$ decreases significantly for the Bell-dominated $\sigma=10^{-3}$ case, due to the lower electron injection rate we see in Figure \ref{fig:acceleration}(b).
The value could become even smaller after longer-term evolution, if the electron injection rate further decreases at later times.

We define the downstream number fraction of the nonthermal particles of species $s$ as:
\begin{equation} \label{eq:nfraction}
\tilde{n}_s = \frac{\int_{\gamma_{\mathrm{inj}}}^{\infty}f_sd\gamma}{\int_1^{\infty}f_sd\gamma}.
\end{equation}
The dotted lines in Figure \ref{fig:energy_partition}(b) show $\tilde{n}_s$.
The values are between $0.04$ and $0.12$ for both ions and electrons, and have a similar $\sigma$-dependence as the energy fractions $\varepsilon_{s}$.

We have also computed the ion-to-electron effective mass ratio:
\begin{equation} \label{eq:eff_mass_ratio}
\left[\frac{m_{\mathrm{i}}}{m_{\mathrm{e}}}\right]_{\mathrm{eff}}=\frac{m_{\mathrm{i}}}{m_{\mathrm{e}}}\frac{\int_1^{\infty}\gamma f_{\mathrm{i}}(\gamma)d\gamma}{\int_1^{\infty}\gamma f_{\mathrm{e}}(\gamma)d\gamma}.
\end{equation}
The values for the three lower-$\sigma$ cases $(\sigma=10^{-4.5}, 10^{-4}$, and $10^{-3.5})$ are all around $[m_{\mathrm{i}}/m_{\mathrm{e}}]_{\mathrm{eff}}\sim5$, while $[m_{\mathrm{i}}/m_{\mathrm{e}}]_{\mathrm{eff}}\sim10$ for the $\sigma=10^{-3}$ case.
This implies that our simulation mass ratio $m_{\mathrm{i}}/m_{\mathrm{e}}=100\gg[m_{\mathrm{i}}/m_{\mathrm{e}}]_{\mathrm{eff}}$ is sufficiently large to capture, both qualitatively and quantitatively, the key aspects of ion and electron energy partitioning.
In other words, due to efficient heating to ultra-relativistic temperatures, electrons lose memory of the initial mass difference.
Previous studies have confirmed this conclusion with mass ratio surveys for ultra-relativistic and trans-relativistic shocks \citep{Spitkovsky2008a,Sironi2013,Crumley2019}.

In this Subsection, we have investigated the particle acceleration physics in different $\sigma$ regimes.
While ions can be accelerated in all cases, efficient electron acceleration is only possible for low-$\sigma$ Weibel-dominated shocks.

\section{Summary and Discussion} \label{sec:discussion}
\subsection{Summary of the Results} \label{subsec:summary}
Here, we summarize our main results.
Let us first summarize the three main results discussed in Section \ref{sec:results}.
Recall that our fixed parameters are: the shock Lorentz factor $\Gamma_{\mathrm{sh}}\sim2$ and the angle $\theta_B=20^{\circ}$ between the shock normal and the ambient mean magnetic field, defined in the upstream frame.
We have investigated the magnetization dependence in the range from $\sigma=10^{-4.5}$ to $\sigma=10^{-3}$.
Our main findings are as follows:
\begin{enumerate}
\item 
The instability that governs the upstream field fluctuations is the Bell instability for a relatively high magnetization, $\sigma\gtrsim10^{-3}$, and the Weibel instability at lower magnetization, $\sigma\lesssim10^{-4}$.
The magnetic field energy density in the near upstream region reaches $\varepsilon_B\sim10^{-2}$ for Bell-dominated cases, and $10^{-4}<\varepsilon_B<10^{-3}$ for  Weibel-dominated cases.
\item 
Ions are efficiently accelerated in both the Bell-dominated and the Weibel-dominated cases, in the sense that their nonthermal energy fraction is $0.1\lesssim\varepsilon_{\mathrm{i}}\lesssim0.2$. 
However, the rate at which the maximum ion energy grows is different between the two regimes.
We find a Bohm-like $E_{\mathrm{max,i}}\propto t$ scaling for the Bell-dominated $\sigma=10^{-3}$ case, and $E_{\mathrm{max,i}}\propto t^{1/2}$ for  Weibel-dominated cases.
\item 
A significant amount of the upstream energy is converted into downstream electron energy.
At lower magnetizations $(\sigma\lesssim10^{-4})$, when the shock is Weibel-mediated, the electrons receive a fraction $0.3\lesssim \mathcal{E}_{\mathrm{e}}\lesssim0.4$ of the upstream flow energy.
In the Bell-dominated regime $\sigma\gtrsim10^{-3}$, this value is reduced to about $\mathcal{E}_{\mathrm{e}}\sim0.2$. 
There is a stark difference in the fraction of upstream flow energy channeled into nonthermal electrons between the two regimes.
Bell-dominated shocks inject only a few percent of the available energy into nonthermal electrons $(\varepsilon_{\mathrm{e}}\ll0.1)$, whereas Weibel-dominated shocks convert a similar amount of energy to nonthermal electrons as to nonthermal ions $(\varepsilon_{\mathrm{e}}\sim\varepsilon_{\mathrm{i}}\sim0.1)$.
\end{enumerate}
We discuss the structure of unmagnetized $(\sigma=0)$ trans-relativistic shocks in Appendix \ref{app:unmagnetized}, details of the high-current Bell modes in Appendix \ref{app:saturation}, and the self-regulation of cosmic ray current in Appendix \ref{app:self_regulation}.

\subsection{Maximum Ion Energy} \label{subsec:max_energy}
Let us discuss the maximum energy of ions that can be attained in trans-relativistic shocks.
We start with relatively high magnetization environments, where a Bohm-like scaling, $E_{\mathrm{max,i}}\propto t$, is applicable.
The magnetic field geometry at termination shocks of astrophysical jets is not well known.
If there exist extended regions with a subluminal shock configuration, then these shocks would be efficient particle accelerators and we assume here for simplicity that the maximum proton energy is constrainted only by the system-size (Hillas) limit as: $E_{\mathrm{max,i}}\sim eBL$, where $B$ and $L$ are the characteristic magnetic field strength and the characteristic size of the system, respectively.
For the termination shock of a typical extragalactic jet,
\begin{equation}
E_{\mathrm{max,i}}\sim10^{20}\left(\frac{B}{10\,\mathrm{\mu G}}\right)\left(
\frac{L}{10\,\mathrm{kpc}}\right)\,\mathrm{eV},
\end{equation}
which makes them promising candidates for UHECR production \citep{Cerutti2023,Globus2025}.
For a microquasar like SS 433,
\begin{equation}
E_{\mathrm{max,i}}\sim10^{16}\left(\frac{B}{10\,\mathrm{\mu G}}\right)\left(
\frac{L}{1\,\mathrm{pc}}\right)\,\mathrm{eV},
\end{equation}
so the recent detection of $\sim100\,\mathrm{TeV}$ gamma rays \citep{Cao2025} could have a hadronic origin.
Bell-dominated trans-relativistic shocks may also be relevant to ion acceleration and gamma ray emission in ultrafast outflows (UFOs) from AGN \citep{Ajello2021,Nishiura2026}.
Applications to the nonthermal emission from BL Lacertae objects have also been discussed \citep{Crumley2019,Arbet-Engles2025}.

For particle acceleration in these relatively highly magnetized shocks, it is essential that the shock velocity is trans-relativistic, instead of ultra-relativistic.
As we have discussed in Section \ref{sec:intro} and \ref{sec:method}, ultra-relativistic shocks are inevitably superluminal.
For any finite magnetization, the maximum energy in ultra-relativistic shocks is limited by the magnetization as $\gamma_{\mathrm{max}}\propto\sigma^{-1/4}$ \citep{Sironi2013, Plotnikov2018}.
Note that, however, \citet{Reville2014} proposed a stronger scaling $(\gamma_{\mathrm{max}}\propto\sigma^{-1/2})$.

\subsection{Synchrotron Emission} \label{subsec:emission}
Here, we discuss thermal and nonthermal synchrotron emission from trans-relativistic astrophysical shocks.
Let us start with the nonthermal emission from late-time GRB afterglows.
Previous studies show that ultra-relativistic low-magnetization shocks can efficiently generate nonthermal electrons \citep{Spitkovsky2008a, Sironi2013, Groselj2024}.
Observations of GW170817 indicate that a significant amount of nonthermal electron energy is required to model the late-time trans-relativistic stage of the GRB external shock \citep{Margutti2018,Hajela2019,Hajela2022}.

However, the PIC simulations by \citet{Crumley2019} imply that the electron nonthermal energy fraction is small, i.e., $\varepsilon_{\mathrm{e}}\ll0.1$.
We have shown that this only holds at relatively high ambient magnetizations $(\sigma\gtrsim10^{-3})$, which are considerably higher than those expected in GRBs.
Instead, a typical interstellar-like medium has a low magnetization:
\begin{equation}
\sigma_{\mathrm{ISM}}=0.5\times10^{-9}\left(\frac{B}{3\,\mathrm{\mu G}}\right)^2\left(\frac{n}{1\,\mathrm{cm}^{-3}}\right)^{-1}.
\end{equation}
At low magnetizations $(\sigma\lesssim10^{-4})$, trans-relativistic shocks become Weibel-dominated and convert a significant amount of shock energy to nonthermal electrons $(\varepsilon_{\mathrm{e}}\sim0.1)$, which is similar to what was found in the ultra-relativistic regime.

Recent studies on FBOTs, such as AT2018cow \citep{Ho2019,Margutti2019}, reveal that synchrotron emission in those systems could instead be dominated by relativistic thermal electrons.
Our results show substantial electron heating at all magnetizations.
Furthermore, we have obtained microphysical parameters $\varepsilon_B, \mathcal{E}_{\mathrm{e}},$ and $\varepsilon_{\mathrm{e}}$, in the trans-relativistic, weakly magnetized regime.
These quantities can be used, for example, in models that infer shock velocity and ambient density from observed light curves \citep{Margalit2021,Margalit2024}.

\subsection{Deceleration Signature} \label{subsec:deceleration}
In this paper, we have studied shocks with Lorentz factor $\Gamma_{\mathrm{sh}}\sim2$.
Here, we shall discuss the physics in different shock velocity regimes and how that can be used to extract information from decelerating shocks.
Here, we refer to the deceleration of astrophysical shocks as they progressively slow down while propagating into their surrounding medium, on dynamical timescales (therefore, much longer than the plasma scales of our simulations).
Time evolution of macroscopic structures, during ultra-relativistic and non-relativistic stages, can be described by the Blandford-McKee solution \citep{Blandford1976} and the classical Taylor-von Neumann-Sedov solution \citep[e.g.,][]{Landau1987}, respectively.

Let us start with the ultra-relativistic limit $(\Gamma_{\mathrm{sh}}\gg1)$.
In this case, the shock will be quasi-perpendicular, regardless of the upstream $\theta_B$.
This regime has been investigated extensively both for electron-positron and electron-ion plasmas \citep{Kato2007,Kato2008,Spitkovsky2008a,Spitkovsky2008b,Sironi2013,Groselj2022}.
The shock is Weibel-dominated, and thus the maximum particle energy has a $E_{\mathrm{max}}\propto t^{1/2}$ scaling.
It is worth pointing out that moderately magnetized $10^{-4}\lesssim\sigma\lesssim10^{-3}$ electron-ion shocks are still relatively under-explored.
In this magnetization range, the Weibel instability may be modified by the ambient field, affecting the electron physics \citep{Jikei2024a}.
In non-relativistic shocks, this can trigger magnetic reconnection near the shock front \citep{Matsumoto2015,Bohdan2021}.
Similar arguments may also apply to relativistic shocks, especially at relatively low Lorentz factors $\Gamma_{\mathrm{sh}}\lesssim10$.

We now move to shock Lorentz factors smaller than $\Gamma_{\mathrm{sh}}\sim2$.
In the case of non-relativistic shocks, we expect a quasi-parallel shock to be Bell-dominated even at very low magnetization, or equivalently, at very high Alfv\'{e}nic Mach number $M_{\mathrm{A}}$.
Hybrid simulations show efficient magnetic amplification by the Bell instability \citep{Caprioli2014b}.
Intense heating of the incoming electrons would be required for efficient magnetic field generation by the Weibel instability \citep{Lyubarsky2006,Jikei2024b}.
Heating of electrons to a level that significantly reduces the effective mass ratio 
$[m_{\mathrm{i}}/m_{\mathrm{e}}]_{\mathrm{eff}}$ is not possible in the non-relativistic regime.
Although the suppression of electron injection by the superluminality constraint does not apply in the non-relativistic limit, the nonthermal electron energy fraction is still small $(\varepsilon_{\mathrm{e}}\ll0.1)$ in non-relativistic Bell-dominated shocks \citep{Park2015}.

These considerations lead us to the speculation that almost every relativistic shock wave starts as a Weibel-dominated shock and eventually transitions to a Bell-dominated state after deceleration to non-relativistic speeds.
Since Weibel-dominated shocks convert significant energy into nonthermal electrons $(\varepsilon_{\mathrm{e}}\sim0.1)$, while $\varepsilon_{\mathrm{e}}$ is much smaller for Bell-dominated shocks, we argue that the nonthermal luminosity should abruptly decrease when the shock decelerates to non-relativistic speeds.
For example, the observation of relatively bright nonthermal emission for GW170817 for more than 1000 days \citep{Hajela2019, Hajela2022} is consistent with the common assumption that the upstream environment is low-$\sigma$ and that the shock speed is still trans-relativistic.
As a direct follow-up to this work, we will characterize the transition (in shock speed) from a Weibel-dominated to a Bell-dominated shock.

\begin{acknowledgments}
T.J. and L.S. are supported by a grant from the Simons Foundation (MP-SCMPS-0000147).
D.G. is supported by the Research Foundation--Flanders (FWO) Senior Postdoctoral Fellowship 12B1424N.
L.S. was also supported by the NSF grant PHY2409223, by the Multimessenger Plasma Physics Center (MPPC, NSF grant PHY2206609), and by the DoE Early Career Award DE-SC0023015. 
This research used computational resources of the National Energy Research Scientific Computing Center (NERSC), a Department of Energy User Facility using NERSC award FES-ERCAP0033914, of the supercomputer of ACCMS, Kyoto University, through the HPCI System Research Project (Project ID: hp250112), and of the Flatiron Institute. 
We acknowledge the OSIRIS Consortium, consisting of UCLA and IST (Lisbon, Portugal), for the use of OSIRIS and for providing access to the OSIRIS framework.
We thank Damiano Caprioli, Raffaella Margutti, Brian D. Metzger, Kaya Mori, and Anatoly Spitkovsky for insightful discussions.
\end{acknowledgments}

\appendix
\section{Unmagnetized Trans-Relativistic Shock} \label{app:unmagnetized}
In the main text, we performed shock simulations in the magnetization range $\sigma=[10^{-4.5},10^{-3}]$.
We refrain from directly comparing to lower magnetizations for the following reason.
Since we are running our simulations up to the same time in units of the plasma time $\omega_{\mathrm{pi}}t$, the final state of the simulations at lower magnetizations are earlier in units of the ion gyration time $\Omega_{\mathrm{i}}t$, where $\Omega_{\mathrm{i}}=eB_0/m_{\mathrm{i}}c$.
For instance, $\omega_{\mathrm{pi}}t=7000$ for $\sigma=10^{-4.5}$ corresponds to $\Omega_{\mathrm{i}}t\sim40$, which is large enough for returning particles to become gyrotropic and drive instabilities with growth rates comparable to the gyro-frequency.
However, this would not hold for even lower magnetizations.
Furthermore, the Larmor radius of particles becomes larger for lower $\sigma$, and we would miss structures on Larmor scales if we used a fixed box size in units of the plasma skin depth.

\begin{figure}[ht!]
\includegraphics[width=\columnwidth]{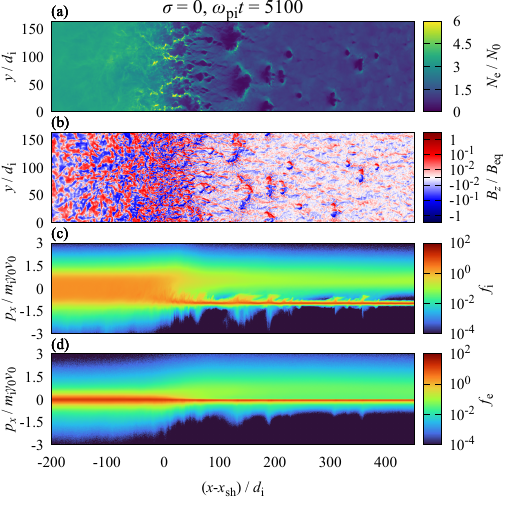}
\caption{Snapshot of the unmagnetized shock with $\sigma=0$, taken at $\omega_{\mathrm{pi}}t=5100$. The format here is mostly the same as in Figure \ref{fig:1e-3}; however, $B_x$ and $B_y$ are not shown since they are always zero here.}
\label{fig:unmag}
\end{figure}

With the above caveats in mind, we present here, for completeness, the unmagnetized case.
Figure \ref{fig:unmag} shows a snapshot of an unmagnetized $\sigma=0$ shock taken at $\omega_{\mathrm{pi}}t=5100$.
The simulation parameters are the same as described in the main text (Section \ref{sec:method}).
In panels (a,b), we see magnetized plasma cavities that are noticeably larger than the ion skin depth scale.
The role of plasma cavities has been extensively discussed in recent works, focusing on the ultra-relativistic regime \citep{Peterson2022,Groselj2024,Demidov2026}.
Here, we find that plasma cavities can also be generated in trans-relativistic unmagnetized shocks.
As found in the ultra-relativistic regime, these cavities may enable efficient field generation and particle acceleration.
The behavior seen in Figure \ref{fig:unmag} may be representative of Weibel-dominated shocks at magnetizations smaller than the range covered in the main text, e.g., $\sigma\sim10^{-9}$, as appropriate for the interstellar medium.

In this Appendix, we have shown the shock structure for the unmagnetized $\sigma=0$ case.
We confirm that the large-scale cavity structure in unmagnetized ultra-relativistic shocks is also present in our trans-relativistic regime.
This implies that particle acceleration in low-$\sigma$ trans-relativistic shocks is not limited by the spatially small structures we see in the early stages of Weibel instability, since at sufficiently low magnetizations, larger structures will be generated at later times.

\section{Saturation Level of the Bell Instability} \label{app:saturation}
Assessing the saturation level of the Bell instability is essential to fully understand the competition with the Weibel instability.
Here, we illustrate the saturation level for both the low-current and the high-current regimes by revisiting previous studies and performing 1D PIC simulations in the parameter space most relevant for our work.

Two scalings have been proposed in the literature regarding the saturation of Bell modes.
The first one pertains to the low-current regime and dictates that the magnetic field energy saturates at a fixed fraction of the cosmic ray momentum flux.
The idea originates from the pioneering work by \citet{Bell2004}.
\citet{Zacharegkas2024} have recently formulated the saturation mechanism in a more precise manner, and validated their predictions by means of hybrid-kinetic simulations.
They propose 
\begin{equation}
\frac{\delta B^2}{8\pi\beta_{\mathrm{d}}T^{01}_{\mathrm{CR}}}\sim\frac{1}{4},
\end{equation}
where $T_{\mathrm{CR}}$ is the cosmic ray stress-energy tensor in the upstream frame, and $\beta_{\mathrm{d}}=v_{\mathrm{d}}/c$ is the drift velocity of the cosmic rays, in the upstream frame.
For an ion-dominated cosmic ray population with an isotropic and mono-energetic distribution in its rest frame, the momentum flux reads:
\begin{equation}
\beta_{\mathrm{d}}T^{01}_{\mathrm{CR}}=m_{\mathrm{i}}c^2n_{\mathrm{CR}}\gamma^2_{\mathrm{d}}\beta^2_{\mathrm{d}}\frac{4\gamma^2_{\mathrm{iso}}-1}{3\gamma_{\mathrm{iso}}}.
\end{equation}
Here, $n_{\mathrm{CR}}$ and $\gamma_{\mathrm{iso}}$ are the number density and the Lorentz factor of the cosmic ray particles, in their rest frame.
Note that we have also defined the drift Lorentz factor $\gamma_{\mathrm{d}}=(1-\beta^2_{\mathrm{d}})^{-1/2}$.

Recent works showed that the nature of Bell modes in the high-current regime $\eta>\eta_{\mathrm{crit}}$ is different \citep{Weidl2019,Lichko2025}.
Here, the magnetic energy density of Bell modes saturates at a fixed fraction of the ambient field energy.
This is common among gyro-resonant plasma instabilities \citep{Stix1992}.
\citet{Lichko2025} found that $\delta B^2/B_0^2\sim25$ for the high-current regime of the Bell instability.

We perform 1D periodic box PIC simulations to clarify the saturation level in the parameter regime of interest, which can be seen as a trans-relativistic generalization of previous studies.
We set up the simulation in the upstream frame with a background ion density of $n_0$.
The ambient field $\bm{B}_0$ is in the $x$-direction, along the simulation box.
Definitions of the background plasma quantities, $\omega_{\mathrm{p}s}, \Omega_s ,d_s$, and $\sigma$ are the same as in the main text.
The cosmic ray component has a proper density of $n_{\mathrm{CR}}$, drifting in the positive $x$-direction with velocity $v_{\mathrm{d}}$.
Note that the simulation frame number density is $\gamma_{\mathrm{d}}n_{\mathrm{CR}}$, and the cosmic ray current (normalized to $en_0c$) is $\eta=\beta_{\mathrm{d}}\gamma_{\mathrm{d}}n_{\mathrm{CR}}/n_0$.
For charge and current neutrality, the background electrons are initialized with a simulation frame density of $n_0+\gamma_{\mathrm{d}}n_{\mathrm{CR}}$ and drift velocity of $v_{\mathrm{d}} \gamma_{\mathrm{d}}n_{\mathrm{d}}/(n_0+\gamma_{\mathrm{d}}n_{\mathrm{CR}})$.
We fix the drift velocity $\beta_{\mathrm{d}}=0.8$, and investigate the dependence on $\eta, \gamma_{\mathrm{iso}}$ and $\sigma$.
The simulation parameters are $\Delta x/d_{\mathrm{e}}=0.1, c\Delta t/\Delta x=0.99$, and $L_x=153.6 d_{\mathrm{i}}$.
We use $N_{\mathrm{ppc}}=1024$ particles per cell, with a quartic spline shape function, for each of the three components, which are background ions, cosmic ray ions, and electrons.
We use the open source code SMILEI \citep{Derouillat2018} in this Appendix.

\begin{figure}[ht!]
\includegraphics[width=\columnwidth]{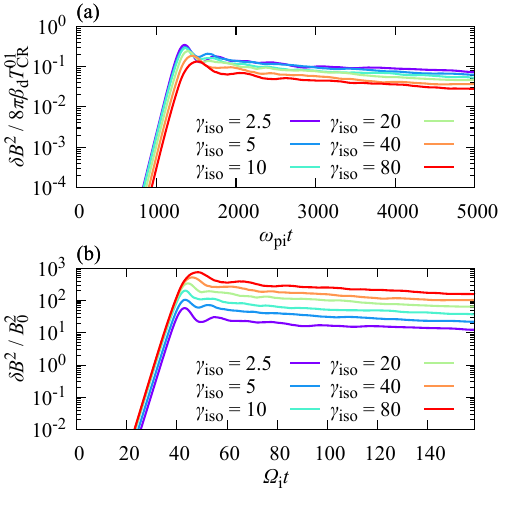}
\caption{Time evolution of the magnetic field energy for the low-current regime. In panel (a), the magnetic field energy is normalized by the cosmic ray momentum flux, and time is normalized by the ion plasma frequency. In panel (b), the normalizations are based on the background magnetic field energy and the ion gyro-frequency, respectively. The colors correspond to different cosmic ray energies. Purple represents the lowest $\gamma_{\mathrm{iso}}=2.5$, and red represents the highest $\gamma_{\mathrm{iso}}=80$.}
\label{fig:low_current}
\end{figure}

Let us start with the low-current regime.
Figure \ref{fig:low_current} shows the results for fixed magnetization $\sigma=10^{-3}$ and current $\eta=0.02$, resulting in $\eta/\eta_{\mathrm{crit}}=0.32$.
We vary the cosmic ray energy $\gamma_{\mathrm{iso}}=[2.5,80]$.
The time evolution of the spatially averaged magnetic field energy $\delta B^2=B_y^2+B_z^2$ is shown with two different normalizations.
Panel (a) uses $8\pi\beta_{\mathrm{d}}T^{01}_{\mathrm{CR}}$ normalization for the magnetic field energy and $\omega_{\mathrm{pi}}$ for the time.
The growth rate is consistent with the theoretical prediction $\Gamma/\omega_{\mathrm{pi}}=\eta/2$ \citep{Bell2004, Amato2009}. 
At the peak $\omega_{\mathrm{pi}}t\sim1500$, we measure the saturation level of $\delta B^2/8\pi\beta_{\mathrm{d}}T^{01}_{\mathrm{CR}}$ in the range of $[0.1,0.4]$, which is consistent with the value $\sim1/4$ by \citet{Zacharegkas2024}.
However, we see a systematic $\gamma_{\mathrm{iso}}$-dependence of $\delta B^2/8\pi\beta_{\mathrm{d}}T_{\mathrm{CR}}^{01}\propto\gamma_{\mathrm{iso}}^{-0.2}$, which was not discussed in their work.
This is most likely due to relativistic effects absent in hybrid simulations.
Panel (b) uses $B^2_0$ and $\Omega_{\mathrm{i}}$ for normalization of magnetic energy and time, respectively.
We can see that the magnetic field amplification factor increases with larger cosmic ray energy, with the scaling $\delta B^2/B_0^2\propto\gamma_{\mathrm{iso}}^{0.8}$.
Again, this is a slightly weaker scaling than $\delta B^2/B_0^2\propto\gamma_{\mathrm{iso}}$ as proposed by \citet{Zacharegkas2024}.
Field amplification proportional to cosmic-ray energy presents favorable conditions for the long-term acceleration of particles at low-current Bell-dominated shocks.

\begin{figure}[ht!]
\includegraphics[width=\columnwidth]{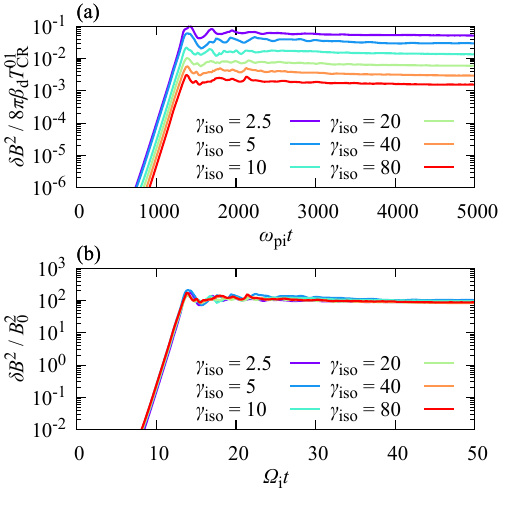}
\caption{Time evolution of the magnetic field energy for the high-current regime. The format and the color scheme are the same as in Figure \ref{fig:low_current}.}
\label{fig:high_current}
\end{figure}

Figure \ref{fig:high_current} shows the results for the high-current regime.
The magnetization and the current are $\sigma=10^{-4}$ and $\eta=0.1$, resulting in $\eta/\eta_{\mathrm{crit}}=5$.
The growth rate is consistent with the high-current Bell theoretical prediction $\Gamma=\Omega_{\mathrm{i}}$ \citep{Weidl2019}.
In panel (b), we can see a very consistent $\delta B^2/B_0^2\simeq100$.
This is comparable with the results by \citet{Lichko2025} with a factor 4 difference.
When normalized by the cosmic ray momentum flux (panel (a)), it is apparent that high-current Bell modes cannot efficiently tap into the cosmic ray energy, especially at high $\gamma_{\mathrm{iso}}$.

\begin{figure}[ht!]
\includegraphics[width=\columnwidth]{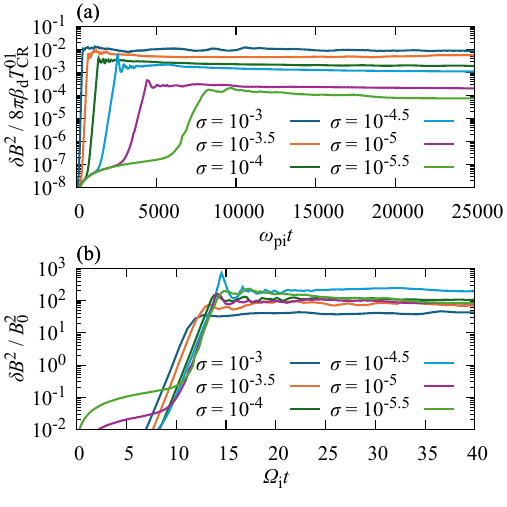}
\caption{Magnetization dependence for the saturation level of the high-current Bell modes. The format is the same as in Figure \ref{fig:low_current}. The color representation for $\sigma$ is the same as the main text (Figures \ref{fig:cr_current} and \ref{fig:eps_b}) for $\sigma=[10^{-4.5},10^{-3}]$, and 2 additional $\sigma=10^{-5}$ (plum) and $10^{-5.5}$ (green) are addded.}
\label{fig:sigma_dependence}
\end{figure}

Figure \ref{fig:high_current} and the results by \citet{Lichko2025} elucidate the saturation level of the high-current Bell instability for a fixed $\sigma$.
To fully understand the transition from a Bell-dominated shock to a Weibel-dominated shock, as we saw in the main text (Subsection \ref{subsec:shock_structure}), we also need to clarify the dependence on the magnetization.
Figure \ref{fig:sigma_dependence} shows the result for fixed $\gamma_{\mathrm{iso}}=10$ and $\eta=0.1$, while varying $\sigma$.
Panel (a), in which the fluctuating field energy is normalized to the cosmic ray momentum flux, shows a clear decline in the saturation level at lower magnetizations.
In panel (b), we see that the growth rate is consistently $\Gamma=\Omega_{\mathrm{i}}$.
For the saturation level, it was expected that the $\sigma=10^{-3}$ case, in dark teal, would perform differently, since $\eta/\eta_{\mathrm{crit}}$ is only slightly above unity, at $1.6$.
At magnetizations below $\sigma=10^{-3}$, which are significantly supercritical $(\eta/\eta_{\mathrm{crit}}\gg1)$, a similar $\delta B^2/B_0^2\sim100$ magnetic field saturation level as in Figure \ref{fig:high_current} is found.
Although we see some variation in the saturation level, we find that there is no systematic $\sigma$-dependence when the instability operates in the high-current regime.

In this Appendix, we studied the saturation level of Bell instability in the regime of trans-relativistic drift velocity via 1D PIC simulations.
In the low-current regime, the magnetic field amplification scales in proportion to the cosmic ray momentum flux.
On the other hand, in the high-current regime, magnetic field amplification is limited by the ambient field strength.
At low magnetizations, it is inevitable that the Weibel instability, whose saturation level is not capped by the ambient field strength, becomes the dominant magnetic field generation mechanism.

\section{Self-Regulation of Cosmic Ray Current and Energy Partition} \label{app:self_regulation}
In the main text (Section \ref{sec:results}), we saw that the strength of the cosmic ray current has a significant impact on shock structure and particle acceleration.
Here, we discuss the different types of shock configurations that can be envisioned in the trans-relativistic regime, as a result of nonlinear feedback between accelerated particles and their self-generated fields.

Let us start by enumerating the possible conditions of the cosmic ray population (see Table \ref{tab:state} for a summary).
First, we have the case in which the current is dominated by ions, and the current is subcritical $(\eta<\eta_{\mathrm{crit}})$.
This, by definition, results in a low-current Bell-dominated system.
Hereafter, we shall call this state I.
Next, we have the case in which the current is dominated by ions, and its magnitude is supercritical $(\eta>\eta_{\mathrm{crit}})$.
As we have discussed in Appendix \ref{app:saturation}, high-current Bell modes would then be dominant for relatively high magnetizations (state IIa), and Weibel modes would win for lower magnetizations (state IIb).
Finally, we have the case where, as a result of efficient electron heating and acceleration, the number and the energy fractions of cosmic ray ions and electrons are comparable, giving a negligible current.
Only Weibel modes can grow in this case, because there is no net current to drive Bell-type instabilities (state III).
We have confirmed the steadiness of state I by running the $\sigma=10^{-3}$ case up to $\omega_{\mathrm{pi}}t=8500$ and of state III by running the $10^{-4}$ case up to $\omega_{\mathrm{pi}}t=12000$ (not shown).

Now we argue that the shock configuration corresponding to state II is generally unstable, meaning that the shock will transition to either state I or III.
Consider a population of shock-reflected electrons with energies comparable to the typical downstream energy, under state IIa.
Their Larmor radius in the amplified field can be written as
\begin{equation}\label{eq:larmor}
\begin{aligned}
\frac{r_{\mathrm{g,e}}}{d_{\mathrm{i}}}=&7\left(\frac{\sigma}{10^{-3}}\right)^{-1/2}\left(\frac{\delta B/B_0}{10}\right)^{-1}\\
&\times\left(\frac{u_{\mathrm{ref}}/c}{7}\right)\left(\frac{[m_{\mathrm{i}}/m_{\mathrm{e}}]_{\mathrm{eff}}}{10}\right)^{-1/2}.
\end{aligned}
\end{equation}

We have made the following assumptions.
We take $\delta B/B_0\sim10$ as a universal value for the high-current Bell regime appropriate for state IIa (Appendix \ref{app:saturation}).
We assume elastic reflection at the shock front, which leads to a typical four-velocity after reflection of $u_{\mathrm{ref}}\sim\Gamma_{\mathrm{sh}}^2V_{\mathrm{sh}}\sim7c$.
The mass ratio $m_{\mathrm{i}}/m_{\mathrm{e}}$ is $\sim1836$ for protons for realistic parameters, and 100 in our PIC simulations.
However, accounting for electron heating, the effective mass ratio becomes $[m_{\mathrm{i}}/m_{\mathrm{e}}]_{\mathrm{eff}}\sim10$ (see Subsection \ref{subsec:acceleration}).
If this Larmor radius is smaller than the typical half wavelength of high-current Bell modes, $\lambda/2d_{\mathrm{i}}\sim10$, which we assume to be independent of the initial $m_{\mathrm{i}}/m_{\mathrm{e}}$ (Figure \ref{fig:1e-3.5}(c)), electron reflection is suppressed.
Note that electrons are always affected more dramatically than ions, due to their smaller Larmor radius.
If this happens, the shock downstream becomes negatively charged, and the electrostatic field pulls back some of the returning ions towards the downstream.
This reduces the cosmic ray current, resulting in the transition to an ion-dominated low-current cosmic ray state, that is, state I.
This is consistent with the time evolution of the $\sigma=10^{-3}$ case in the main text.
The dark teal curve in Figure \ref{fig:cr_current} reaches a high-current state at $\omega_{\mathrm{pi}}t\sim2000$, but self-regulates to a low-current state at $\omega_{\mathrm{pi}}t\gtrsim5000$.

On the other hand, at lower magnetizations $\sigma\lesssim10^{-4}$, the electron Larmor radius in the saturated high-current Bell field is larger than the half wavelength.
Therefore, electrons will be more efficiently reflected back into the upstream.
As a result, the system transitions to state III (ion and electron cosmic rays).
The same can be said for state IIb, since Weibel modes do not prevent electron reflection, also resulting in efficient electron acceleration.

In this Appendix, we have categorized the different cosmic ray conditions and argued that high-current Bell-dominated shock configurations are generally unstable (meaning, a transient state) and could transition to a low-current Bell or Weibel-dominated system, depending on the magnetization.
This argument indicates that our results in the main text are not early-time transients.
In addition, it also elucidates the transition from the low-current Bell-dominated regime at $\sigma=10^{-3}$ to the Weibel-dominated regime at $\sigma=10^{-4}$, as observed in our simulations.
A high-current Bell-dominated system was present at $\sigma=10^{-3.5}$, but this case may not have reached the steady state at $\omega_{\mathrm{pi}}t=7000$.
We have run the $\sigma=10^{-3.5}$ simulation for a longer time, up to $\omega_{\mathrm{pi}}t=10000$, and the late stage (not shown) shows a reduction in electron injection, which may drive the system towards the low-current Bell-dominated state I.
Table \ref{tab:state} summarizes the above argument.

\onecolumngrid
\begin{table*}[ht!]
\begin{tabular}{l|llll}
State     & CR composition   & CR current   & Magnetic field     & Steady State?         \\ \hline
I         & Ion-dominated    & Low-current  & Low-current Bell   & Yes                  \\
IIa       & Ion-dominated    & High-current & High-current Bell  & No: to I $(\sigma\gtrsim10^{-3})$ or III $(\sigma\lesssim10^{-4})$ \\
IIb       & Ion-dominated    & High-current & Weibel             & No: to III      \\
III       & Ion and Electron & None         & Weibel             & Yes             
\end{tabular}
\caption{Summary of the different types of shock configurations in Appendix \ref{app:self_regulation}. We consider four states (I, IIa, IIb, and III), based on the differences in cosmic ray characteristics and magnetic field structures. States I and III can be steady states of shock upstream. State IIa can transition to state I or III, depending on the magnetization. IIb always transitions to state III.}
\label{tab:state}
\end{table*}

\bibliography{main}{}
\bibliographystyle{aasjournalv7}

\end{CJK*}
\end{document}